\documentclass[aps,prl,floatfix,twocolumn,groupedaddress]{revtex4}

\usepackage{graphicx}
\usepackage{epstopdf}
\usepackage{times}
\usepackage{amssymb,amsmath}
\usepackage{array}
\usepackage{multirow}

\newcommand{\ket}[1]{\left\vert#1\right\rangle}

\def\degree{\ensuremath{{}^{\circ}}}

\begin{document}
\title{Experimental parity-time symmetry quantum walks on a directed graph}

%\author
\author{Tong Wu$^{1}$, J. A. Izaac$^{2}$, Zi-Xi Li$^{1}$, Kai Wang$^{1}$, Zhao-Zhong Chen$^{1}$, Shining Zhu$^{1}$, J. B. Wang$^{2}$, Xiao-Song Ma$^{1,\ast}$}
\affiliation{
$^1$~National Laboratory of Solid-state Microstructures, School of Physics, Collaborative Innovation Center of Advanced Microstructures, Nanjing University, Nanjing 210093, China\\
$^2$~School of Physics, The University of Western Australia, Crawley, WA 6009, Australia\\
$^{\ast}$e-mail:Xiaosong.Ma@nju.edu.cn}

\date{\today}

\begin{abstract}
	Quantum walks (QW) are of crucial importance in the development of quantum information processing algorithms. Recently, several quantum algorithms have been proposed to implement network analysis, in particular to rank the centrality of nodes in networks represented by graphs. Employing QW in centrality ranking is advantageous comparing to certain widely used classical algorithms (e.g. PageRank) because QW approach can lift the vertex rank degeneracy in certain graphs. However, it is challenging to implement a directed graph via QW, since it corresponds to a non-Hermitian Hamiltonian and thus cannot be accomplished by conventional QW. Here we report the realizations of centrality rankings of both a three-vertex and four-vertex directed graphs with parity-time (PT) symmetric quantum walks. To achieve this, we use high-dimensional photonic quantum states, optical circuitries consisting of multiple concatenated interferometers and dimension dependent loss. Importantly, we demonstrate the advantage of QW approach experimentally by breaking the vertex rank degeneracy in a four-vertex graph. Our work shows that PT-symmetric quantum walks may be useful for realizing advanced algorithm in a quantum network.
\end{abstract}
\maketitle
\onecolumngrid

\section*{Introduction}
Classical random walks are of fundamental importance in a variety of scientific disciplines, including physics, biology, and economics. Quantum walks, first introduced by Aharonov, Davidovich and Zagury~\cite{Aharonov:1993to}, are the quantum analogue of classical random walks~\cite{Kempe:2003xg}. Quantum walks harness the superposition and the entanglement of quantum systems. These unique features of quantum walks lead to a number of applications in the field of quantum information theory, including the development of quantum-enhanced algorithms~\cite{Ambainis:2003sj, Shenvi:2003el,Szegedy:2004cf,Berry:2010qp}, quantum simulations~\cite{Mulken:2011ta, Berry:2011lc, Ahlbrecht:2012dn,Izaac:2017cd,Izaac:2017lq} and universal quantum computation~\cite{Childs:2009oq,Childs:2013xr}. Quantum walks have been realized in a variety of physical systems, including photons~\cite{Bouwmeester:1999lz,Du:2003ef,Schreiber:2010vk,Broome:2010ai,Peruzzo:2010qg,Schreiber:2012xm,Kitagawa:2012ql,Jeong:2013hj,Qiang:2016uu,Xiao:2017te,Tang:2018ws,Qiang:2018aa}, and natural and artificial atoms~\cite{Schmitz:2009dm,Karski:2009qf,Zahringer:2010sw,Preiss:2015zc,Yan:2019zc}. See review in~Ref. \onlinecite{Manouchehri:2014ws}.

One promising application of quantum walks is to provide an efficient quantum algorithm for vertex centrality ranking in network analysis~\cite{Berry:2010qp, Paparo:2012nb,Paparo:2013si,Sinayskiy:2013gd,Loke:2016mw,Falloon:2017me}. Using continuous-time quantum walks (CTQWs), the aforementioned algorithm has been physically implemented on an undirected graph~\cite{Izaac:2017cd}. However, in most applications one needs to perform the analysis on directed graphs, which are more general than undirected graphs in representing practical systems. For instance, in the classical PageRank algorithms, revealing the relevance between websites and listing their ranks require network analysis on directed graphs. It is challenging to rank the centrality of a directed graph with QW experimentally, which stems from the intrinsic conflict between the directionality of the PageRank algorithm and the non-directionality/unitarity of traditional quantum walks. This issue has been theoretically solved by introducing pseudo-Hermitian evolutions~\cite{Bender:1998ik,Salimi:2010pt,Izaac:2017lq}. In this work, we demonstrate the first realization of quantum walks on directed graphs with single photons, and their applications in centrality ranking.

\section*{Quantum walk on a three-vertex directed graph and its experimental realization}

\begin{figure*}[t]
	\includegraphics[width=0.84\textwidth]{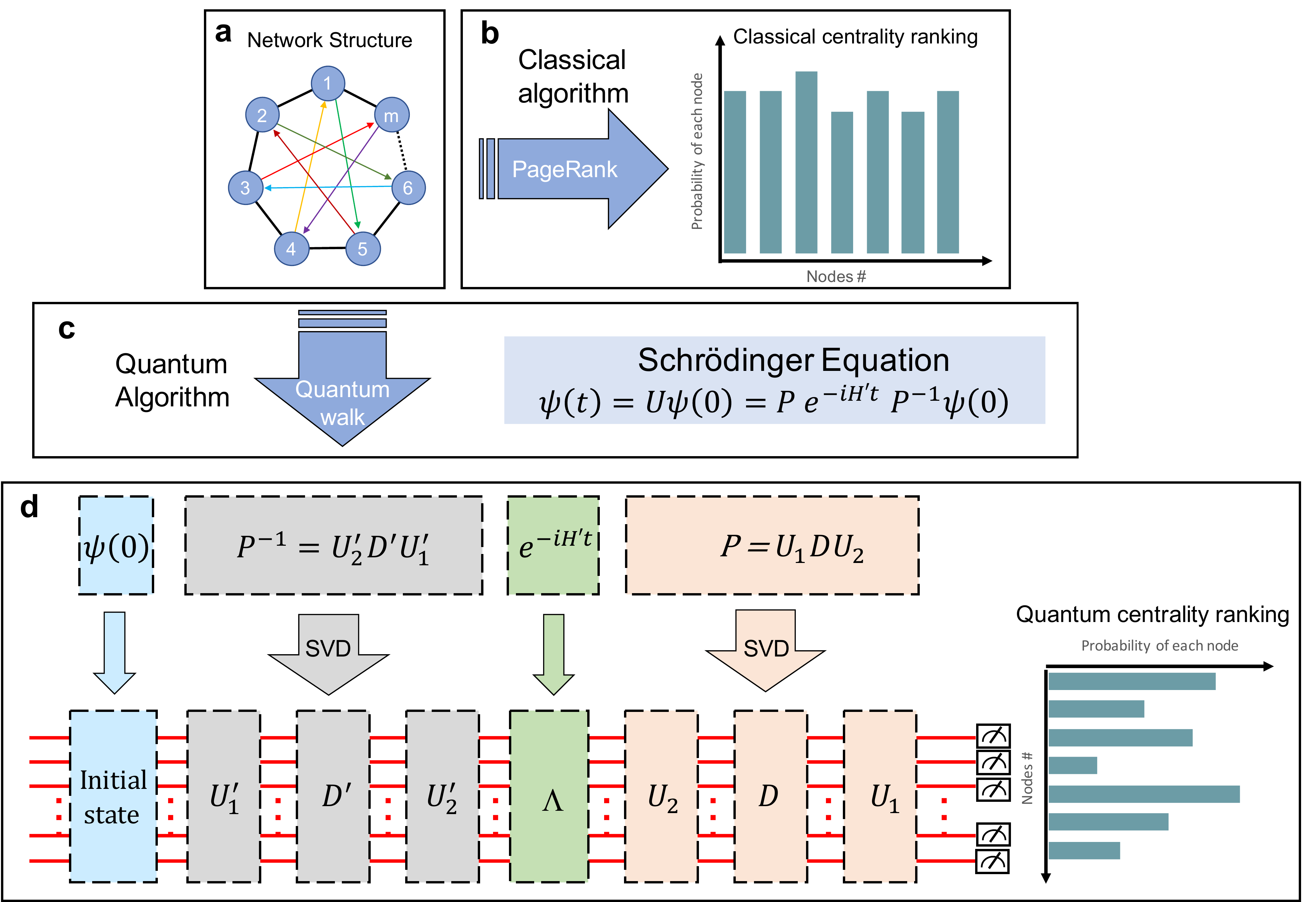}
	\caption{\label{fig:pagerank}Network analysis with classical and quantum algorithms. \textbf{a}. A general network structure can be represented as a directed graph with weighted edges, which are characterized by its adjacency matrix. \textbf{b}. By using classical PageRank algorithm, one can calculate centrality ranking of the graph representing the network. \textbf{c}. In the quantum scenario, one can map the centrality ranking problem to solve a quantum walk dynamics with Schr\"odinger equation. The dynamics is governed by time evolution operator, \textit{U}, which is related to the adjacency matrix of the graph (see text). By diagonalizing the hamiltonian, one decomposes \textit{U} into three parts, the diagonal \textit{H'} and two matrices \textit{P} and \textit{P}$^{-1}$ for a general graph. \textbf{d}. The recipe for performing centrality ranking of a graph with quantum walk. We first generate the initial state ($\ket{\psi(0)}$), whose dimensionality corresponds to the number of vertex in the graph, represented by the red paths. Then we use singular value decomposition (SVD) to decompose \textit{P}$^{-1}$ and \textit{P} into $P^{-1}=U_{2}'D'U_{1}'$ and $P=U_{1}DU_{2}$. The evolution of diagonalized Hamiltonian \textit{H'} can be realized with matrix $\Lambda$. Finally we measure the output from this circuitry and obtained the centrality of the vertex of the graph. }
\end{figure*}

As shown in Fig. 1\textbf{a}, a network can be represented as a graph $G(V,E)$ consisting of vertices $j\in V$ and edges $(i,j)\in E$. The adjacency matrix \textit{A} is used to characterize the graph $G$, defined as
\begin{align}\label{eq:ajacency}
A_{ij} = \begin{cases}
1, & (i,j)\in E\\
0, & (i,j)\notin E ~.
\end{cases}
\end{align}
In a directed (undirected) graph, the adjacency matrix \textit{A} is unsymmetric (symmetric) where $A_{ij}\neq A_{ji}$ ($A_{ij}=A_{ji}$). By using classical algorithm, such as Pagerank, one can calculate centrality ranking of the graph representing the network, as shown in Fig. 1\textbf{b}. On the other hand, one can use quantum algorithm, such as quantum walk, to rank the centrality of each vertex, as shown in Fig.1\textbf{c}. Particularly, in a continuous-time quantum walk on graph $G$, the conjugate transpose of the Hamiltonian is the graph Laplacian of the adjacency matrix and can be written as~\cite{Aharonov:1993to, Kempe:2003xg, Manouchehri:2014ws}:
\begin{align}\label{eq:hamiltonian}
(H^{\dagger})_{ij} = \mathcal{L}_{ij} = \left(\sum_k A_{ik}\right)\delta_{ij} -A_{ij}~.
\end{align}
By solving the Schr\"odinger equation, one obtains the time evolution of the walker and its wave-function, at time \textit{t}, is given by $\ket{\psi(t)} = e^{-iHt}\ket{\psi(0)}$. If the graph is undirected, the adjacency matrix \textit{A} is Hermitian and hence matrix \textit{H} is also Hermitian based on Eq.~(2). In a continuous-time quantum walk on an undirected graph the time evolution operator, $U(t)=e^{-iHt}$, is unitary and the norm of $\ket{\psi(t)}$ is conserved. This is the traditional scenario of continuous-time quantum walks and has been employed in different experimental contexts~\cite{Peruzzo:2010qg, Qiang:2016uu,Izaac:2017cd,Tang:2018ws}.

%On the contrary,
A walker behaves differently on a directed graph, where the adjacency matrix $A$ is no longer symmetric and hence the Hamiltonian $H$ is not Hermitian. Consequently, the time evolution operator is not unitary and the squared norm of $\ket{\psi(t)}$ could grow or decay exponentially.  Note that there have been proposals focusing on the non-unitary evolution, for example, Szegedy quantum walks~\cite{Paparo:2013si,Loke:2017so,Qiang:2018aa} and open-quantum walks~\cite{Sinayskiy:2013gd,Loke:2016mw,Izaac:2017lq}. Here we follow the theoretical model proposed by Izaac et. al.~\cite{Izaac:2017lq} and realize experimentally a quantum walk on a directed graph. In Fig. 1\textbf{d}, we show the recipe for performing centrality ranking of a general graph with quantum walk. Firstly, we generate a high-dimensional initial state, $\ket{\psi(0)}$. Its dimensionality corresponds to the number of vertex in the graph of interest. The Hamiltonian, \textit{H}, can be diagonalized and decomposed into three matrices: \textit{P}$^{-1}$, \textit{H'} and \textit{P}. If the graph is an undirected one, all three matrices are unitary. If not, both \textit{P}$^{-1}$, \textit{P} are not unitary. Moreover, if the eigenvalues of \textit{H} are all real and the eigenvectors are linearly independent, $H$ is pseudo-Hermitian~\cite{Bender:2010hz}. This is the focus throughout this manuscript. We may then use the singular value decomposition (SVD) to decompose \textit{P}$^{-1}$ and \textit{P} into $P^{-1}=U_{2}'D'U_{1}'$ and $P=U_{1}DU_{2}$, where $D'$ ($D$) is a diagonal matrix with non-negative real numbers on the diagonal, which are known as the singular values of $P^{-1}$ ($P$). The evolution of diagonalized Hamiltonian \textit{H'} can be realized with matrix $\Lambda$, which represents the unitary dynamics. Finally we measure the output from this circuitry and obtained the centrality of the vertex of the graph.

In this article, we use two simple yet illustrative exemplary graphs to demonstrate the above mentioned concept. The first one is a three-vertex directed graph shown in Fig. 2\textbf{a}, whose adjacency matrix and the corresponding Hamiltonian are:
\begin{align}\label{eq:3vertexmatrix}
A_{3} = \left[\begin{matrix}
0 &  1 & 0\\
1 &  0 & 0\\
1 &  1 & 0
\end{matrix}\right], \hspace{1cm}  H_{3} = \left[\begin{matrix}
1 & -1 & -1\\
-1 & 1 & -1\\
0 & 0 & 2
\end{matrix}\right]
\end{align}
We can easily verify that $H_{3}$ is pseudo-Hermitian~\cite{Bender:2010hz}.  Although the time evolution is not unitary and hence the total probability is not conserved, the pseudo-Hermitian character of the system ensures that the total probability of the quantum walker's state will just oscillate, with no exponential growth and decay.

We decompose the time evolution operator such that the time dependence is restricted to a single diagonal unitary matrix. We can diagonalize $U$ and obtain the time-evolution operator as:
\begin{align}
U & =  P \Lambda P^{-1} \\
& = \frac{1}{2}\left[\begin{matrix}
-1 & 1 & -1\\
1 & 1 &0 \\
0 & 0 & 1
\end{matrix}\right]\left[\begin{matrix}
e^{-2it} & 0 & 0\\
0 & 1 & 0\\
0 & 0 & e^{-2it}
\end{matrix}\right]\left[\begin{matrix}
-1 & 1 & -1\\
1 & 1 & 1\\
0 & 0 & 2
\end{matrix}\right].
\end{align}
The diagonal matrix $\Lambda$ is unitary and represents the phase accumulation in CTQW, which can be implemented easily as shown in Supplemental Material \uppercase\expandafter{\romannumeral3}B~\cite{SMQRW2019}. However, the two outer matrices, $P$ and $P^{-1}$, are not unitary, and are non-trivial to realize experimentally with linear optics.  We overcome this challenge by expanding the dimensionality of states that the single photon occupies, as well as introducing non-unitary loss. Firstly, we employ a ququart, whose Hilbert space is four-dimensional and is spanned by its horizontal/vertical polarization and two optical paths ($\ket{1H}$,$\ket{1V}$,$\ket{2H}$,$\ket{2V}$). Secondly, we use dimension-dependent photon-number loss. Finally, we verify the non-unitary evolution for three out of four dimensions by obtaining the probability for each vertex.

\begin{figure*}[ht]
	\includegraphics[width=0.9\textwidth]{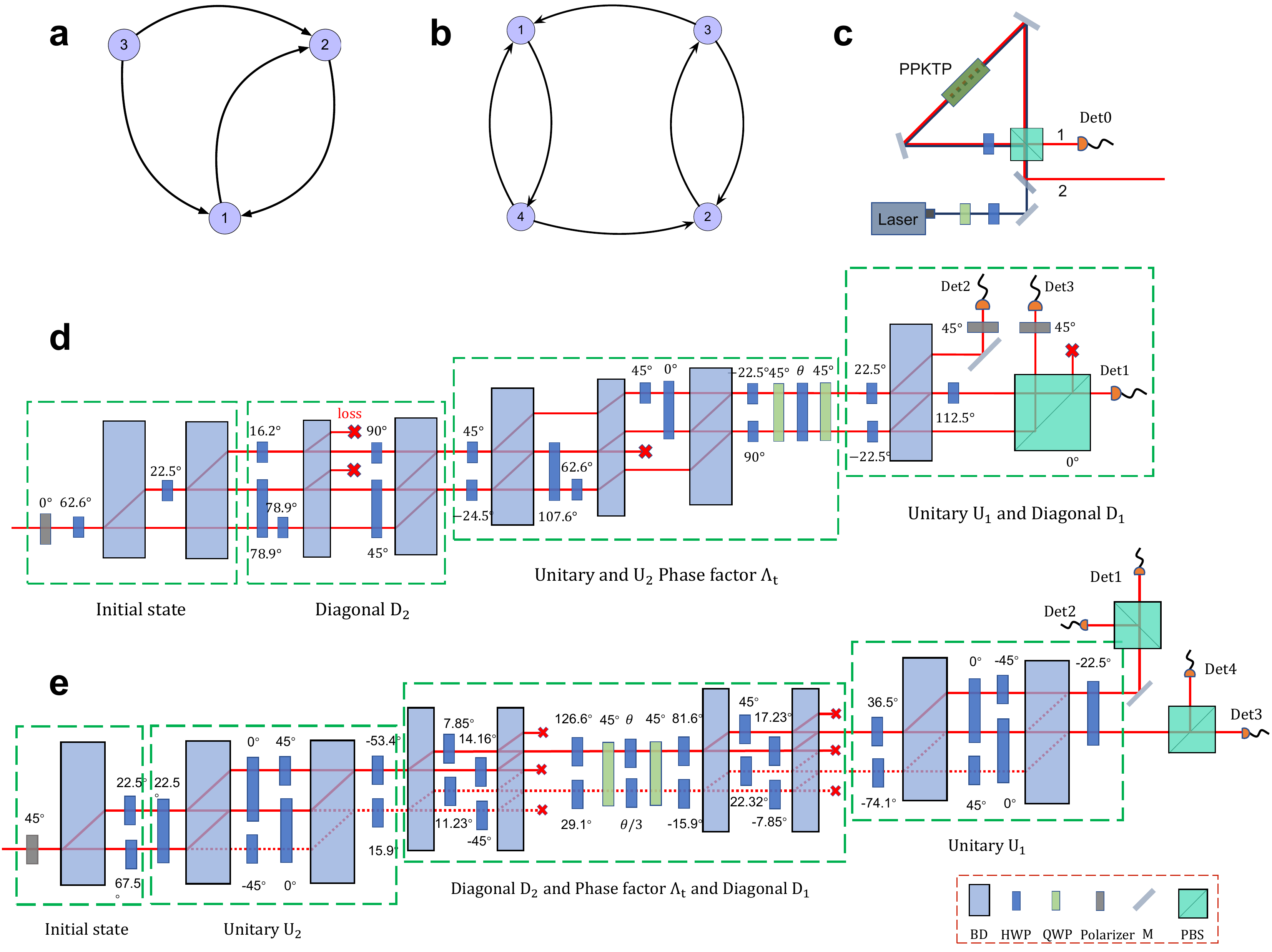}
	\caption{\label{fig:setup}Experimental setup for realizing quantum walk on directed graphs with single photons. A three-vertex, \textbf{a}, and a four-vertex, \textbf{b}, directed graphs are realized in our experiment.  \textbf{c}. The single-photon source. A pair of photons (photons 1 and 2) are generated from spontaneous parametric down conversion process, which is realized by embedding a PPKTP crystal in a Sagnac interferometer. We couple the photons into single-mode fibers, respectively. Photon 1 is detected by Det0, which is a single-photon avalanche diode (SPAD). The single-photon count rate of photon 1 is about 600~kHz. This detection heralds the presence of photon 2. \textbf{d} and \textbf{e} show the optical circuitries for realizing the three-vertex and four-vertex directed quantum walk. To realize the continuous quantum walk on a directed graph, photon 2 is sent through an optical network consisting of beam displacers (BDs), polarizing beam splitter (PBS), half-wave plates (HWPs), quarter-wave plates (QWPs), polarizers (Pols) and mirrors (Ms). This optical network can be divided into several blocks, corresponding to diagonal matrix $D_2$, unitary matrix $U_2$, evolutionary matrix $\Lambda_{t}$, unitary matrix $U_1$ and diagonal matrix $D_1$. The optical transmissions of the three-vertex circuitry and the four-vertex circuitry are about 1.5$\%$ to 4.6$\%$ and  1.9$\%$ to 2.5$\%$, respectively.}
\end{figure*}

Here we elaborate on how the non-unitary matrix $P$ is realized. Its inverse $P^{-1}$ could be realized in a similar manner, which is shown in Supplemental Material \uppercase\expandafter{\romannumeral4}B~\cite{SMQRW2019}. Firstly, we extend the three-dimensional non-unitary matrix $P$ to a four-dimensional unitary matrix, in which the parameters are chosen for the convenience of experimental implementations. We then perform singular value decomposition on $P$, where $P=U_{1}DU_{2}$. The matrix $P$ expansion and decomposition are shown:
\begin{align}
\left[\begin{matrix}
-1 & 1 & -1\\
1 & 1 & 0\\
0 & 0 & 1
\end{matrix}\right]  \rightarrow \left[\begin{matrix}
-1 & 1 & -1 & 1\\
1 & 1 & 0 & 0\\
0 & 0 & 1 & 1\\
-1 & 1 & 1 & -1
\end{matrix}\right]
= 2\left[\begin{matrix}
1 & 0 & 0 & 0\\
0 & \frac{1}{\sqrt{2}} & 0 & 0\\
0 & 0 & \frac{1}{\sqrt{2}} & 0\\
0 & 0 & 0 & 1
\end{matrix}\right]\left[\begin{matrix}
-\frac{1}{2} & \frac{1}{2} & -\frac{1}{2} & \frac{1}{2}\\
\frac{1}{\sqrt{2}} & \frac{1}{\sqrt{2}} & 0 & 0\\
0 & 0 & \frac{1}{\sqrt{2}} & \frac{1}{\sqrt{2}}\\
-\frac{1}{2} & \frac{1}{2} & \frac{1}{2} & -\frac{1}{2}
\end{matrix}\right],
\end{align}
where matrix $U_{1}$ is the identity and is not shown. Based on singular value decomposition, $D$ is a diagonal matrix, which could be realized with mode-dependent photon attenuation by tuning the relevant parameters. By using the cosine-sine decomposition~\cite{Stewart:1982vc,Dhand:2015jt,Izaac:2017cd}, we decompose the matrix $U_{2}$ into four two-dimensional unitary matrices to be conveniently realized with linear optics (see Supplemental Material \uppercase\expandafter{\romannumeral4}C~\cite{SMQRW2019} for details).

\begin{figure*}[ht]
	\includegraphics[width=0.75\textwidth]{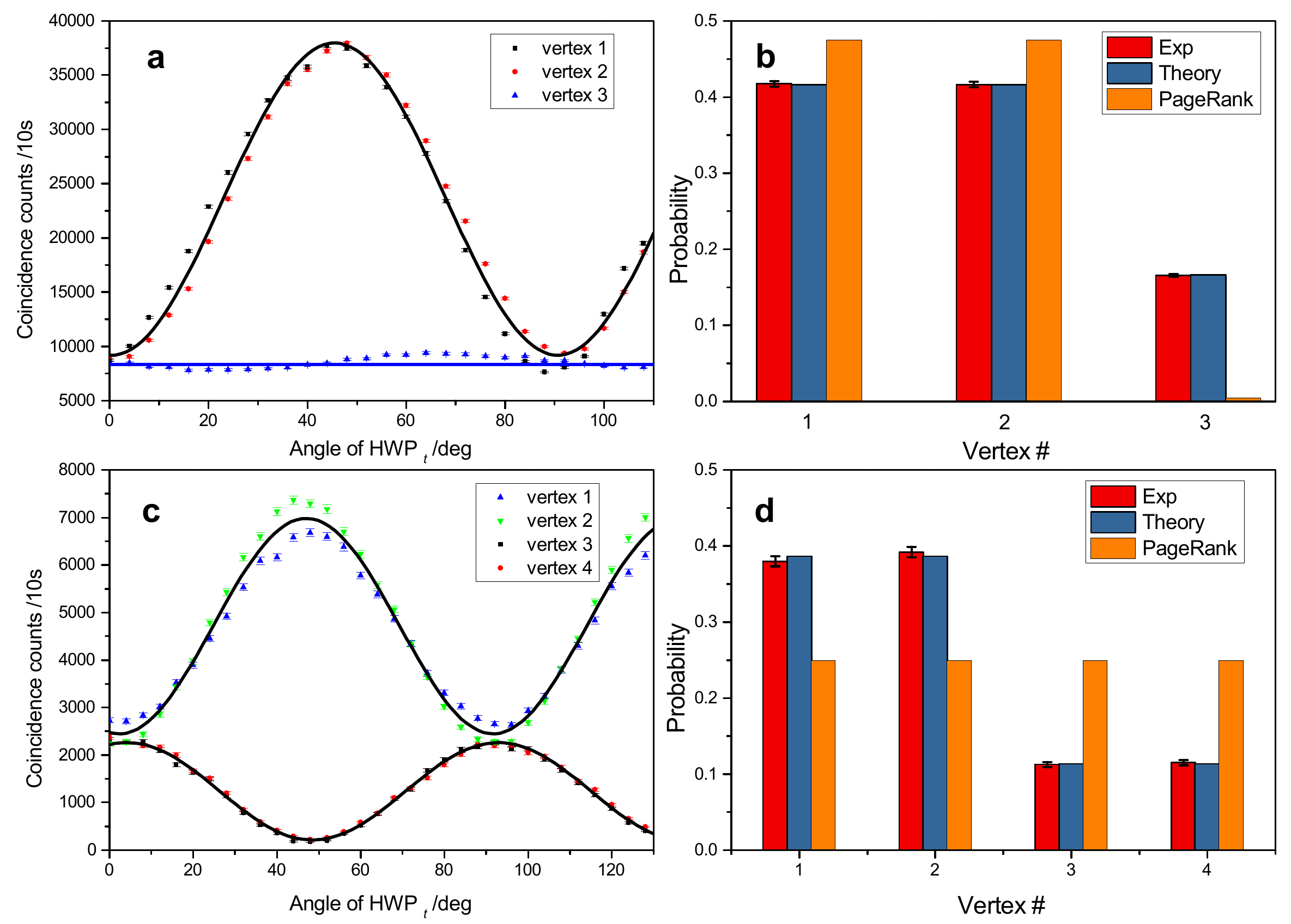}
	\caption{\label{fig:data}Experimental data. \textbf{a}. The black, red and blue data points are the experimental coincidence counts between Det0\&Det1, Det0\&Det2 and Det0\&Det3 as a function of the rotation angle of HWP$_{t}$, respectively. The black curve is the average of the fits to black and red data points. \textbf{b}. The red and blue bars represent the probabilities of photon 2 occupying each vertex averaging over a 2$\pi$ period, which correspond to the centrality ranking of vertex 1, 2 and 3 shown in Fig. 2\textbf{a}. The orange bars indicate the centrality calculated using classical PageRank algorithm. \textbf{c}. The blue, green, black and red data points are the experimental coincidence counts between Det0\&Det1, Det0\&Det2, Det0\&Det3 and Det0\&Det4 as a function of the rotation angle of HWP$_{t}$, respectively. The two black curves are the average of the fits to blue\&green and black\&red data points, respectively. \textbf{d}. The centrality ranking of vertex 1-4 in Fig.2 \textbf{b}. The uncertainties represent standard deviations deduced from propagated Poissonian statistics.}
\end{figure*}

In Fig. 2, we show our experimental schematic of realizing quantum walks on directed graphs. We create pairs of single photons by type-II spontaneous parametric down-conversion in a 20 mm-long periodically-poled potassium titanyl phosphate (PPKTP) crystal, embedded in a Sagnac interferometer~\cite{Kim:2006pi,Fedrizzi:2007bq}. This crystal is pumped by a 10 mW diode laser centered at 405 nm and emits orthogonally polarized photon pairs with a central wavelength of 810 nm and a full-width at the half maximum bandwidth of about 0.5 nm. The 405~nm laser is set to vertical polarization, so that only one arm of the Sagnac interferometer is pumped. As shown in Fig. 2\textbf{c}, in order to create a single-photon Fock state, we detect photon 1 by Det0, which serves as a trigger and heralds the existence of photon 2. We realize the CTQW on the three-vertex directed graph as shown in Fig. 2\textbf{a} with a single-photon ququart.

Based on the process decompositions discussed above, we employ an optical circuitry consisting of beam displacers (BDs), half-wave plates (HWPs) and quarter-wave plates (QWPs) to manipulate the path and polarization states of photon 2 (shown in Fig. 2\textbf{d}). Due to the stability of the concatenated interferometer made by BDs, after an initial phase calibration with classical light, we are able to obtain good results over a few hours without active feedback control. The details of realizing diagonal matrices ($D_1$ and $D_2$) and unitary matrices ($U_1$ and $U_2$) can be found in the Supplemental Material \uppercase\expandafter{\romannumeral1} and \uppercase\expandafter{\romannumeral2}B~\cite{SMQRW2019}. Note that we have stopped the photon occupying mode $\ket{2V}$ after the dimensionality expansion in the step of evolution matrix $\Lambda_{t} $ and $D_1$, respectively. The coincidence detection events of photons at detector Det0 and Det1,  Det0 and Det2, Det0  and Det3 (all with a 2 ns coincidence time window) indicate successful runs of the directed quantum walk, and correspond to centrality probabilities of vertices 1, 2 and 3, respectively. We show the raw data of these three coincidence counts as a function of the angle of HWP$_{t}$ in Fig. 3\textbf{a}. By rotating this angle, we can continuously adjust the phase between the horizontal and vertical polarization states of photon 2. This birefringent phase corresponds to the evolution time of the system, as shown in Eq. (6). It is clear to see that the coincidence counts of Det0\&Det1 and Det0\&Det2  depend strongly  on the angle of HWP$_{t}$.  By contrast, the coincidence counts of Det0\&Det3 is much less sensitive to the angle of HWP$_{t}$. By taking the average values of these three coincidence counts and normalizing them to the sum of these three average values, we obtain the experimental centralities of each vertex respectively (shown in Fig. 3\textbf{b}).

\section*{Quantum walk on a four-vertex directed graph and its experimental realization}
With the three-vertex graph example, we have shown that centrality ranking based on CTQW can reproduce the results derived from classical PageRank algorithm. However, CTQW based algorithm is more advantageous for certain graphs comparing to the classical PageRank algorithm in terms of lifting the degeneracy of centralities of vertices. Surprisingly, this quantum advantage can be already found in a small-scale graph with four vertices, as shown in Fig. 2\textbf{b}. In this case, vertices 1 and 2, and vertices 3 and 4 are two sets of equivalent vertices. However, if we use the classical PageRank algorithm to analyze this graph, we will find that all four vertices have same centralities, which are incorrect. On the other hand, if we use the CTQW, we can break this four-vertex degeneracy and achieve the accurate centralities ranking of vertexes. Particularly, we will show that, by using CTQW, the centralities of vertexes 1 and 2 are higher than that of vertices 3 and 4.

For the four-vertex directed graph shown in Fig. 2\textbf{b}, its adjacency matrix and Hamiltonian can be obtained using Eq.~(1) and Eq.~(2), so:
\begin{align}
\label{eq:3vertex}
A_{4} = \left[\begin{matrix}
0 &  0 & 0& 1\\
0 &  0 & 1& 0\\
1 &  1 & 0& 0\\
1 & 1 & 0 & 0
\end{matrix}\right], \hspace{1cm}  H_{4} = \left[\begin{matrix}
1 & 0& -1 & -1\\
0&1 & -1 & -1\\
0 &-1& 2 & 0\\
-1&0&0&2
\end{matrix}\right]
\end{align}
We can diagonalize its time-evolution operator $U(t)=e^{-iH_d t}$ by $U(t) =  P \Lambda P^{-1} $ for implementing CTQW as:
\begin{align}
U(t) = \frac{1}{6} \left[ \begin{matrix}
2 & -1 & 1 & 0 \\
2 & -1 & -1 & 0 \\
1 & 1 & -1 & 1 \\
1 & 1 & 1 & -1\end{matrix}\right] \left[\begin{matrix}
1 & 0 & 0 & 0 \\
0 & e^{-3it} & 0 & 0 \\
0 & 0 & e^{-it} & 0 \\
0 & 0 & 0 & e^{-2it}\end{matrix}\right] \left[ \begin{matrix}
1 & 1 & 1 & 1 \\
-1 & -1 & 2 & 2 \\
3 & -3 & 0 & 0 \\
3 & -3 & 3 & -3\end{matrix}\right].
\end{align}
With singular value decomposition and cosine-sine decomposition, these matrices can be implemented conveniently using linear optics in a similar manner (see Supplemental Material~\cite{SMQRW2019} for details). In Fig. 2\textbf{e}, we show our experimental setup for realizing quantum walks on the 4-vertex directed graph. Photon 2 is sent through the optical circuitry consisting of beam displacers (BDs), half-wave plates (HWPs) and quarter-wave plates (QWPs) to manipulate the path and polarization states, with its initial state equal to $\ket{1111}$. By solving the time-evolution equation, the photons travel in the solid lines, and there is no photon passing through the dotted lines. We show the experimental data of the four coincidence counts in Fig. 3\textbf{c} and \textbf{d}. The results of the CTQW centrality test agree well with theory, ranking vertices 1 and 2 above vertices 3 and 4.

\section*{Conclusion}
In this study, we have experimentally realized a quantum walk on directed graphs with photonic Fock states. The required non-unitary evolution of the walker on the directed graph is decomposed into a series of unitary and non-unitary transformations in high-dimensional Hilbert spaces spanned by the polarization and spatial modes of photons. We successfully demonstrated quantum centrality rankings on a three-vertex and a nine-vertex graph with PT-symmetric quantum walks, and the advantage of the QW approach by breaking the vertex rank degeneracy in the four-vertex graph comparing with classical PageRank algorithm. Our work shows that, in conjunction with unitary evolutions, controllable loss can be employed as a crucial degree of freedom in quantum walks and may be extended to other quantum information tasks.

\textit{Acknowledgement} The authors thank R. Cao for help during the early stage of this project and X. Jiang for helpful discussions. This research is supported by the National Key Research and Development Program of China (2017YFA0303700, 2019YFA0308704), National Natural Science Foundation of China (Grant No. 11674170, 11621091, 11690032), NSF Jiangsu province (No. BK20170010), the program for Innovative Talents and Entrepreneur in Jiangsu, and the Fundamental Research Funds for the Central Universities.
%\newpage

%\bibliography{bibfileqrw.bib}

%%%%%%%%%% Merge with supplemental materials %%%%%%%%%%
\clearpage \widetext
\begin{center}
	\section{Supplemental Material for ``Experimental parity-time symmetry quantum walks on a directed graph"}
\end{center}
%%%%%%%%%% Prefix a "S" to all equations, figures, tables and reset the counter %%%%%%%%%%
\setcounter{equation}{0} \setcounter{figure}{0}
\setcounter{table}{0} \setcounter{page}{1} \setcounter{secnumdepth}{3} \makeatletter
\renewcommand{\theequation}{S\arabic{equation}}
\renewcommand{\thefigure}{S\arabic{figure}}
\renewcommand{\bibnumfmt}[1]{[S#1]}
\renewcommand{\citenumfont}[1]{S#1}
%\renewcommand\thesection{S\arabic{section}}
%%%%%%%%%% Prefix a "S" to all equations, figures, tables and reset the counter %%%%%%%%%%

\makeatletter
\def\@hangfrom@section#1#2#3{\@hangfrom{#1#2#3}}
\makeatother

%---------------------------------------------------------------------------

\maketitle
\section{Realizations of an arbitrary four-dimensional unitary matrix}\label{sec:one}
It is well known that any discrete finite-dimensional unitary operations could be constructed with optical devices~\cite{Reck:1994ii}. Specifically, here we employ the so-called cosine-sine decomposition (CSD) algorithm~\cite{Stewart:1982vc,Stewart:1977yd,Sutton:2009ei,Dhand:2015jt,Izaac:2017cd} in our study of realizing an arbitrary four-dimensional unitary matrix.

\subsection{Cosine-Sine decomposition (CSD)}\label{sec:onea}
By utilizing the CSD, an arbitrary $(m+n)\times(m+n)$ unitary matrix can be decomposed into a few lower-dimensional unitary matrices and one $S$ matrix of the form:
\begin{align}
U_{m+n} = \mathbb{L}_{m+n}(\mathbb{S}_{2m} \oplus \mathbb{I}_{n-m})\mathbb{R}_{m+n},
\end{align}
where $\mathbb{L}_{m+n}$ and $\mathbb{R}_{m+n}$ are block-diagonal matrices:
\begin{align}
\mathbb{L}_{m+n} = \left[\begin{array}{c|c}
L_{m} &  0  \\
\hline
0 &  L'_{n} \\
\end{array}\right]  ,   \mathbb{R}_{m+n} = \left[\begin{array}{c|c}
R^\dagger_{m} & 0 \\
\hline
0 & R'^\dagger_{n} \\
\end{array}\right] ,
\end{align}
and $\mathbb{S}_{2m}$ is:
\begin{align}
\mathbb{S}_{2m}  & \equiv \mathbb{S}_{2m}(\theta_1,...,\theta_{m})
= \left[\begin{array}{ccc|ccc}
\cos\theta_1 &   &   & \sin\theta_1 &  &  \\
& \ddots & & &\ddots & \cr
&   & \cos\theta_{m} & & & \sin\theta_{m} \\
\hline
-\sin\theta_1 & &    &   \cos\theta_1 & & \\
& \ddots & & & \ddots & \\
&   &  -\sin\theta_{m}  & & & \cos\theta_{m} \\
\end{array}\right]
\end{align}

In the case of a four-dimensional unitary matrix $U_4$, we can decompose it into 4 block matrices via CSD. We rewrite $U_4$ as a block matrix:
\begin{align}\label{eq:ABCD}
U_{4}U^\dagger_4 &= \left[\begin{array}{c|c}
A & B \\
\hline
C & D \\
\end{array}\right] \left[\begin{array}{c|c}
A^\dagger & C^\dagger \\
\hline
B^\dagger & D^\dagger \\
\end{array}\right]
= \left[\begin{array}{c|c}
AA^\dagger + BB^\dagger & AC^\dagger + BD^\dagger \\
\hline
CA^\dagger + DB^\dagger & CC^\dagger + DD^\dagger \\
\end{array}\right]  = I,
\end{align}
where $A$, $B$, $C$, $D$ are all $2\times2$ matrices. Therefore, we obtain:
\begin{align}\label{eq:aabb}
AA^\dagger + BB^\dagger = I \\
CC^\dagger + DD^\dagger = I
\end{align}
Based on algebraic derivations, we can further obtain $[AA^\dagger , BB^\dagger] = [CC^\dagger , DD^\dagger]=0$ and hence $A$ and $B$ ($C$ and $D$) have the same left singular vectors, denoted by $L$ ($L'$). Similarly, from $U^\dagger_4 U_4 = I$, we can obtain $[A^\dagger A , C^\dagger C] = [B^\dagger B, D^\dagger D] = 0$, so $A$ and $C$ ($B$ and $D$) have the same right singular vectors, denoted by $R$ ($R'$). By decomposing $A$, $B$, $C$, $D$ respectively, we can rewrite them as follows:
\begin{align*}
A &= LD_1R\\
B &= LD_2 R' \\
C &= L'D_3 R \\
D &= L'D_4 R',
\end{align*}
where $L$, $L'$, $R$, $R'$ are all two-dimensional unitary matrices and $D_1$, $D_2$, $D_3$, $D_4$ are all two-dimensional, real diagonal matrices. Therefore, $U_4$ can be written as:
\begin{align}\label{eq:U4}
U_4 = L_4S_4R_4 = \left[\begin{array}{c|c}
L &  \\
\hline
&  L' \\
\end{array}\right] \left[\begin{array}{c|c}
D_1 & D_2 \\
\hline
D_3 & D_4 \\
\end{array}\right] \left[\begin{array}{c|c}
R &  \\
\hline
&  R' \\
\end{array}\right]
\end{align}

Since $L$, $L'$, $R$, $R'$, $U_4$ are all unitary matrices and $D_1$, $D_2$, $D_3$, $D_4$ are all two-dimensional, real diagonal matrices, $S_4$ is also a unitary matrix and can be written as:
\begin{align}
S_4 =  \left[\begin{matrix}\label{eq:S4}
\cos\theta_1 & 0 & \sin\theta_1 & 0\\
0 & \cos\theta_2 & 0 & \sin\theta_2\\
-\sin\theta_1 & 0 & \cos\theta_1 & 0\\
0 & -\sin\theta_2 & 0 & \cos\theta_2
\end{matrix}\right].
\end{align}

\subsection{Realization of two-dimensional unitary matrices and the $S_4$ matrix}
According to the above CSD decomposition, for any four-dimensional unitary matrix, we can always decompose it into four two-dimensional unitary matrices and one $S_4$ matrix. In order to realize these matrices, we use the path and polarization modes of lights to encode them, and the base states are $\ket{1H}$, $\ket{1V}$, $\ket{2H}$ and $\ket{2V}$, where the numbers represent the path states, and H and V represent the horizontal and vertical polarization states.

\begin{figure}[htp]
	\includegraphics[scale=0.7]{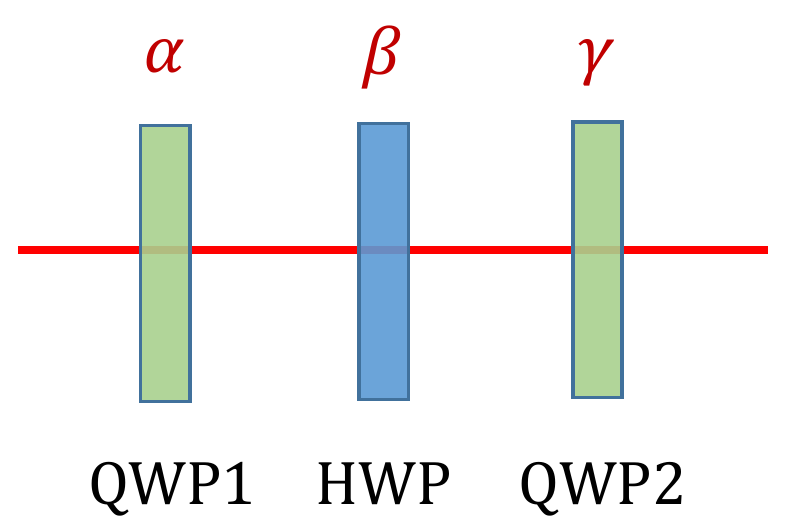}
	\caption{By using a combination of two quarter-wave plates (QWP) and one half-wave plate (HWP), one can achieve arbitrary unitary manipulation of a single photonic qubit encoded in polarization states.}
	\label{fig:HV}
\end{figure}
\begin{figure*}[htp]
	\includegraphics[scale=0.45]{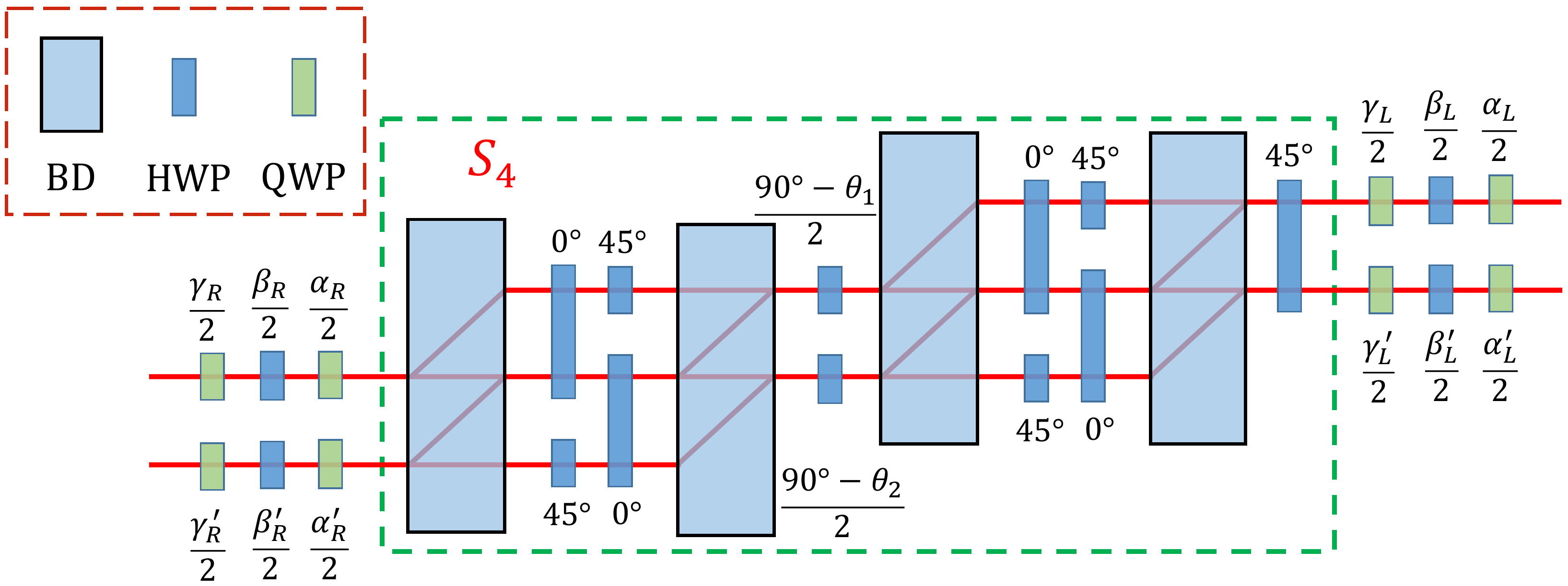}
	\caption{By using a combination of several beam displacers and half-wave plates in the dotted box, we can achieve the $S_4$ matrix. Based on the realization of the $S_4$ matrix and the two-dimensional unitary matrices, we can realize an arbitrary four-dimensional unitary matrix.}
	\label{fig:s_4}
\end{figure*}

A two-dimensional unitary matrix can be written as:
\begin{align}
U_2 & = \left[\begin{matrix}
e^{i\phi}\cos\epsilon & e^{i\xi}\sin\epsilon \\
e^{-i\xi}\sin\epsilon & -e^{-i\phi}\cos\epsilon
\end{matrix}\right],
\end{align}
and we can implement it with wave plates for polarization encoded photonic qubits, as shown in Fig. S1. The angles $\alpha$, $\beta$, $\gamma$ can be obtained by solving the equation based on the Jones matrix of a half-wave plate (HWP) and two quarter-wave plates (QWP)~\cite{Fowles:1989aa}.

To realize the more challenging part, the $S_4$ matrix, we use several beam displacers (BDs) and HWPs, as shown in the dashed box of Fig. S2. Based on the realization of the $S_4$ matrix and the two-dimensional unitary matrices shown in Fig. S1, we can implement an arbitrary four-dimensional unitary matrix, as shown in Fig. S2.

\section{Realizations of non-unitary four-dimensional matrix}\label{sec:two}

In Sec. \uppercase\expandafter{\romannumeral1}, we have introduced the method of implementing an arbitrary four-dimensional unitary matrix via linear optics. However, the case of a non-unitary matrix is fundamentally different from a unitary matrix: the number of photons is no longer conserved. Here we use singular value decomposition (SVD) to facilitate the realization of an arbitrary four-dimensional matrix.

\subsection{Singular value decomposition}\label{sec:twoa}

A matrix can be decomposed into two unitary matrices and a diagonal matrix via SVD. Let $P$ be an arbitrary four-dimensional matrix, and its singular value decomposition has the following form:
\begin{align*}
P &= U_1 D U_2,
\end{align*}
where $U_1$ and $U_2$ are both unitary matrices and $D$ is a real diagonal matrix. $U_1$ and $U_2$ can be realized by the above mentioned method of implementing an arbitrary $4\times4$ unitary matrix. At this point, we only need to realize a real diagonal matrix $D$.

\subsection{Realization of real diagonal matrix}

%which could be realized via mode-dependent photon attenuation. In order to implement $D$, firstly we rewrite $D$ as the form:

$D$ is a real four-dimensional diagonal matrix and can be written as:

\begin{align}\label{eq:diagonal}
D = \lambda\left[\begin{matrix}
\sin2\theta_1 & 0 & 0 & 0\\
0& -\cos2\theta_2 &0 &0\\
0& 0& \sin2\theta_3 &0 \\
0& 0&0 & -\cos2\theta_4
\end{matrix}\right]
\end{align}

We can realize it by tuning the parameters $\theta_1, \theta_2, \theta_3, \theta_4$ as shown in Fig. S3. The beams are separated by BD, then the attenuation of each beam passing through the polarizer is adjusted by changing the angle of the HWPs. By controlling the attenuation of each beam, we can realize the matrix $D$ via dimension-dependent photon attenuation.
\begin{figure}[htp]
	\includegraphics[scale=0.5]{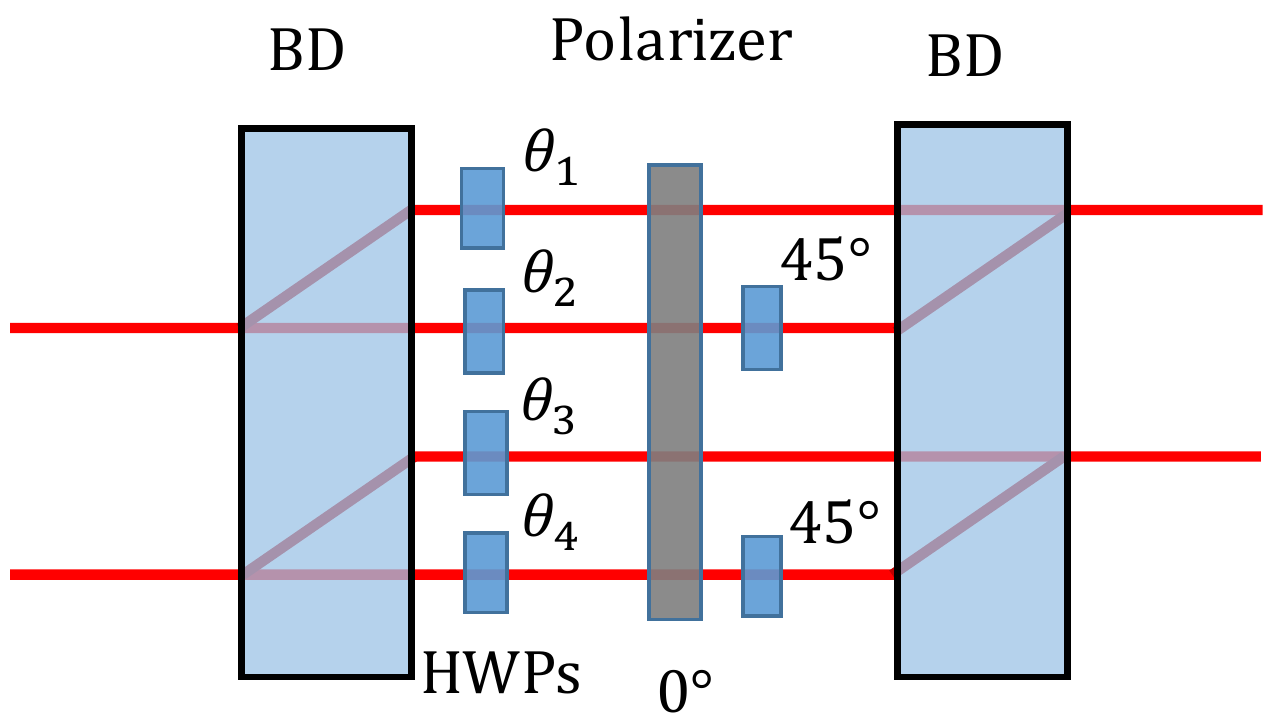}
	\caption{Realization of a real diagonal matrix. We can achieve the real diagonal matrix by controlling the attenuation of each beam by HWPs and a polarizer oriented along 0\degree.}
	\label{fig:diagonal}
\end{figure}

With the realization of a real diagonal matrix shown in Fig. S3 and two four-dimensional unitary matrices shown in Fig. S2, we can implement an arbitrary non-unitary four-dimensional matrix.

\section{Realizations of four-dimensional pseudo-Hermitian quantum walks}

The Hamiltonian in our experiment denoted by $H$ are two four-dimensional pseudo-Hermitian matrices that we will discuss in Sec. \uppercase\expandafter{\romannumeral4}\textbf{B} and Sec. \uppercase\expandafter{\romannumeral5}\textbf{A}. We need to achieve the time evolution matrix $e^{-iHt}$ and the initial state $\ket{\psi(0)}$. $\ket{\psi(0)}$ is easy to implement by using polarization optical elements. In this section we will introduce the method to implement $U(t)=e^{-iHt}$ using linear optical elements.

\subsection{Diagonalization of Hamiltonian}

In order to achieve $U(t)$, firstly we diagonalize the Hamiltonian and rewrite it as $H=P \Lambda P^{-1}$, and then obtain the phase factor for the continuous-time quantum walk (CTQW):
\begin{align}
U(t)=Pe^{-i\Lambda t}P^{-1},
\end{align}
where $\Lambda$ is a diagonal matrix and $P$ and $P^{-1}$ can be treated as two arbitrary four-dimensional matrices. We can implement them according to Sec. \uppercase\expandafter{\romannumeral2}. Next we have to implement the phase factor matrix $e^{-i\Lambda t}$.

\subsection{Realization of the phase factor matrix}
Since $\Lambda$ is a real diagonal matrix, the diagonal matrix $e^{-i\Lambda t}$ is unitary and represents the phase accumulation in CTQW, which can be implemented with the setup as shown in Fig. S4.
\begin{figure}[htp]
	\includegraphics[scale=0.7]{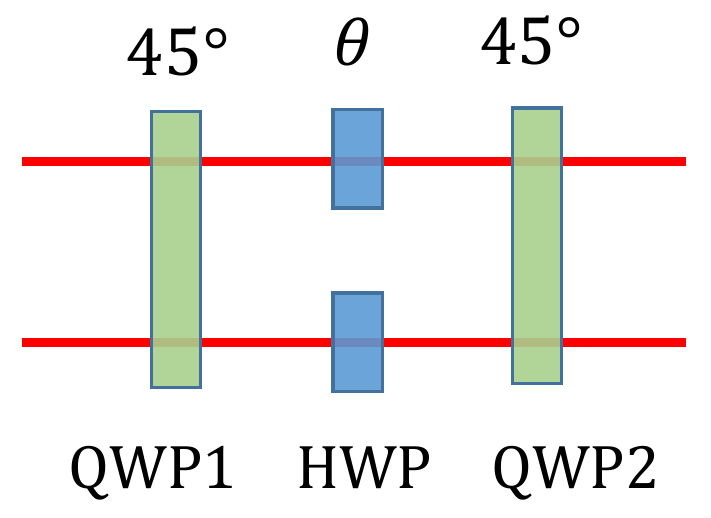}
	\caption{By using a combination of two quarter-wave plates (oriented to $45^\circ$) and one half-wave plate (oriented to $\theta$), we can realize the $\theta$-dependent phase factor.}
	\label{fig:phase}
\end{figure}

\subsection{Schematics of the pseudo-Hermitian continuous-time quantum walks}
Based on the method introduced in Sec. \uppercase\expandafter{\romannumeral2}\textbf{A}, we perform SVD on $P$ and $P^{-1}$ respectively:
\begin{align*}
P&=U_1 D U_2 ,\\
P^{-1}&=U_2^{'} D^{'} U_1^{'}.
\end{align*}
Then we perform CSD on $U_1$, $U_2$, $U_1^{'}$ and $U_2^{'}$ respectively, as introduced in Sec. \uppercase\expandafter{\romannumeral1}\textbf{A}, and after that we can design the setup of a corresponding time evolution matrix, based on the above methods of realizing these matrices. In conjunction with the input state initialization, we can obtain the setup for arbitrary four-dimensional pseudo-Hermitian CTQW, as shown in Fig. S5.
\begin{figure}[htp]
	\includegraphics[width=0.8\textwidth]{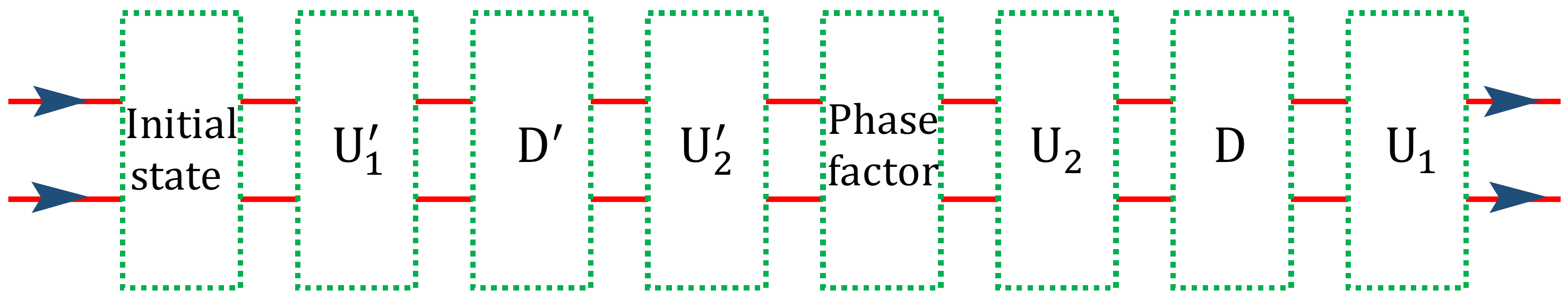}
	\caption{\label{fig:M}Realization of an arbitrary four-dimensional pseudo-Hermitian continuous-time quantum walk}
\end{figure}

\section{Details of implementing the three-dimensional quantum walk }
In this section, we show the details on realizing the pseudo-Hermitian CTQW on our three-vertex directed graph.

\subsection{Diagonalization of Hamiltonian}\label{sec:4a}
As shown in Eq.(4) in the main text, the Hamiltonian of our three-vertex graph is a non-Hermitian matrix. In order to simplify the experimental setup, we chose one of the possible diagonalized matrices, as we will explain later in Sec. \uppercase\expandafter{\romannumeral4}\textbf{B} and Sec. \uppercase\expandafter{\romannumeral4}\textbf{C}. We have
\begin{align}
H & =\left[\begin{matrix}
1 & -1 & -1\\
-1 & 1 & -1\\
0 & 0 & 2
\end{matrix}\right] = \frac{1}{2}\left[\begin{matrix}
-1 & 1 & -1\\
1 & 1 & 0\\
0 & 0 & 1
\end{matrix}\right]\left[\begin{matrix}
2 & 0 & 0\\
0 & 0 & 0\\
0 & 0 & 2
\end{matrix}\right]\left[\begin{matrix}
-1 & 1 & -1\\
1 & 1 & 1\\
0 & 0 & 2
\end{matrix}\right]=P \Lambda P^{-1},
\end{align}
and then we can obtain the time evolution matrix:
\begin{align}
e^{-iHt}  = \frac{1}{2}\left[\begin{matrix}
-1 & 1 & -1\\
1 & 1 & 0\\
0 & 0 & 1
\end{matrix}\right]\left[\begin{matrix}
e^{-2it} & 0 & 0\\
0 & 1 & 0\\
0 & 0 & e^{-2it}
\end{matrix}\right]\left[\begin{matrix}
-1 & 1 & -1\\
1 & 1 & 1\\
0 & 0 & 2
\end{matrix}\right],
\end{align}
where the phase factor matrix (middle) can be achieved with the optical path of Fig. S4. Therefore, we focus on the realization of the following matrices:
\begin{align*}
P = \left[\begin{matrix}
-1 & 1 & -1\\
1 & 1 & 0\\
0 & 0 & 1
\end{matrix}\right] \quad \textrm{and} \quad
P^{-1} = \left[\begin{matrix}
-1 & 1 & -1\\
1 & 1 & 1\\
0 & 0 & 2
\end{matrix}\right]
\end{align*}

\subsection{Dimension expansion and Singular value decomposition}\label{sec:4b}
We perform SVD on the dimension-expanded matrix, and then perform CSD on the unitary matrices. In order to facilitate the experiment we let the CSD produce an identity matrix, so that we can have fewer unitary matrices that need to be realized. Therefore, we expand the matrices that meet these conditions in the following:
\begin{align*}
&P= \left[\begin{matrix}
-1 & 1 & -1\\
1 & 1 & 0\\
0 & 0 & 1
\end{matrix}\right]  \rightarrow \left[\begin{matrix}
-1 & 1 & -1 & 1\\
1 & 1 & 0 & 0\\
0 & 0 & 1 & 1\\
-1 & 1 & 1 & -1
\end{matrix}\right]  = 2\left[\begin{matrix}
1 & 0 & 0 & 0\\
0 & \frac{1}{\sqrt{2}} & 0 & 0\\
0 & 0 & \frac{1}{\sqrt{2}} & 0\\
0 & 0 & 0 & 1
\end{matrix}\right]\left[\begin{matrix}
-\frac{1}{2} & \frac{1}{2} & -\frac{1}{2} & \frac{1}{2}\\
\frac{1}{\sqrt{2}} & \frac{1}{\sqrt{2}} & 0 & 0\\
0 & 0 & \frac{1}{\sqrt{2}} & \frac{1}{\sqrt{2}}\\
-\frac{1}{2} & \frac{1}{2} & \frac{1}{2} & -\frac{1}{2}
\end{matrix}\right]=D_1 U_1
\end{align*}
\begin{align*}
P^{-1}=\left[\begin{matrix}
-1 & 1 & -1\\
1 & 1 & 1\\
0 & 0 & 2
\end{matrix}\right]  \rightarrow \left[\begin{matrix}
-1 & 1 & -1 & \frac{2}{\sqrt{3}}\\
1 & 1 & 1 & -\frac{2}{\sqrt{3}}\\
0 & 0 & 2 & \sqrt{3}\\
-2 & 0 & 1 & -\frac{2}{\sqrt{3}}
\end{matrix}\right] = \sqrt{7}\left[\begin{matrix}
-\frac{1}{\sqrt{6}} & \frac{1}{\sqrt{2}} & -\frac{1}{\sqrt{7}} & \frac{2}{\sqrt{21}}\\
\frac{1}{\sqrt{6}} & \frac{1}{\sqrt{2}} & \frac{1}{\sqrt{7}} & -\frac{2}{\sqrt{21}}\\
0 & 0 & \frac{2}{\sqrt{7}} & \frac{3}{\sqrt{21}}\\
-\frac{2}{\sqrt{6}} & 0 & \frac{1}{\sqrt{7}} & -\frac{2}{\sqrt{21}}
\end{matrix}\right]\left[\begin{matrix}
\sqrt{\frac{6}{7}} & 0 & 0 & 0\\
0 & \sqrt{\frac{2}{7}} & 0 & 0\\
0 & 0 & 1 & 0\\
0 & 0 & 0 & 1
\end{matrix}\right]=U_2 D_2
\end{align*}

The diagonal matrices $D_1$ and $D_2$ can be realized by the optical setup shown in Fig. S3, and the only thing we need is to calculate the angles of those HWPs. We will introduce the method of implementing the unitary matrices $U_1$ and $U_2$ in Sec. \uppercase\expandafter{\romannumeral4}\textbf{C}.

\subsection{Cosine-Sine decomposition of $U_1$ and $U_2$}\label{sec:4c}
If one of the parameters of $S_4$ in Eq. S8 is 0, the setup for realizing $S_4$ will become easier. Let $\theta_1=\theta, \theta_2=0$, then the matrix $S_4$ becomes $S_4'$ :
\begin{align*}
S_4' = \left[\begin{matrix}
\cos\theta & 0 & \sin\theta & 0\\
0 & 1 & 0 & 0\\
-\sin\theta & 0 & \cos\theta & 0\\
0 & 0 & 0 & 1
\end{matrix}\right]
\end{align*}

Compared to the setup of $S_4$ in Fig. S2, the setup of $S_4'$ is simplfied as shown in Fig. S6. To realize the $S_4'$ matrix, it requires a careful choice when performing the diagonalization of the Hamiltonian $H$ in Sec. \uppercase\expandafter{\romannumeral4}\textbf{A} and the expansion of $P$ and $P^{-1}$ in Sec. \uppercase\expandafter{\romannumeral4}\textbf{B}:

\begin{figure}[htp]
	\includegraphics[width=0.6\textwidth]{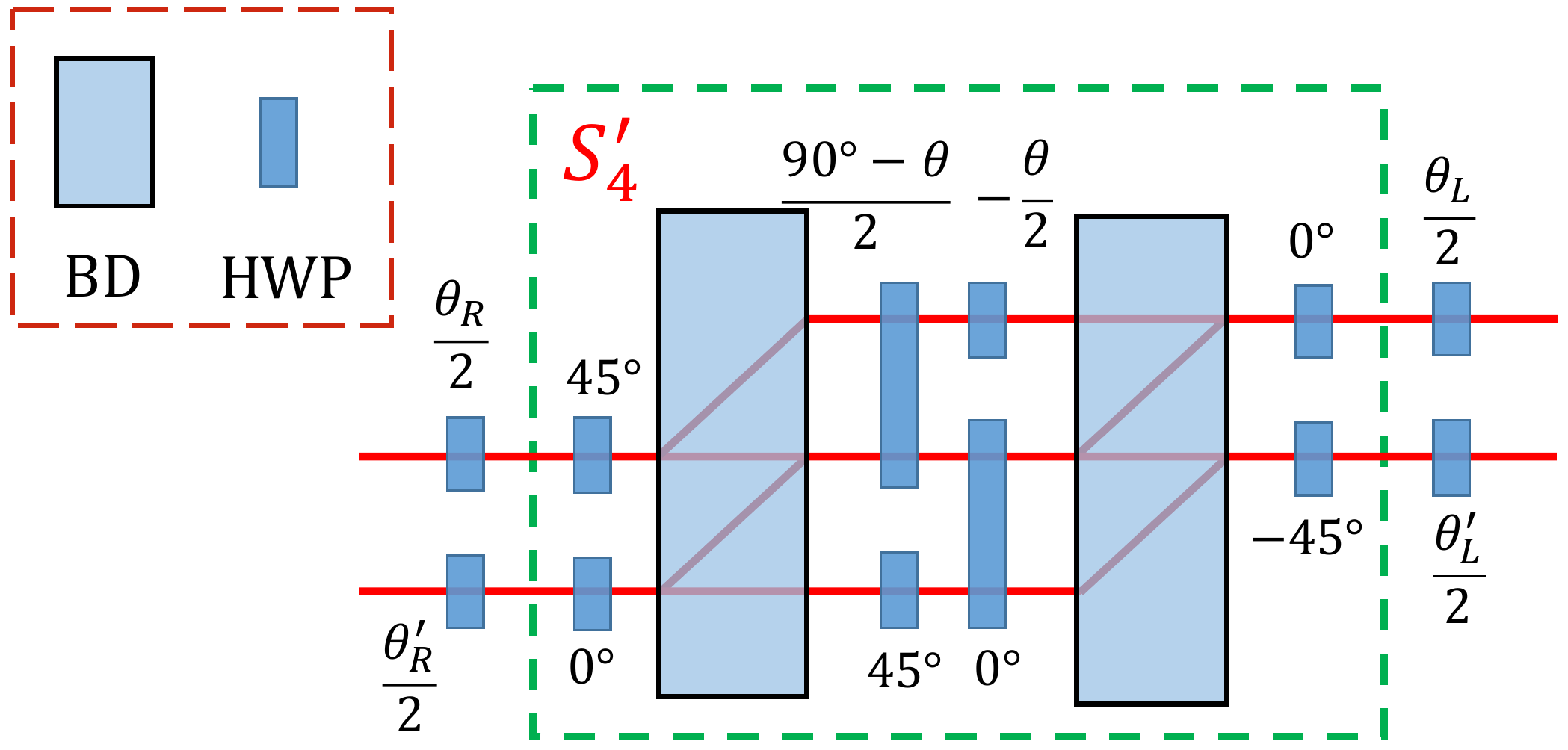}
	\caption{\label{fig:S_4_0}Simplified realization of the four-dimensional unitary matrix. The two-dimensional unitary matrices in our case are all real matrices, only four HWPs are needed. Therefore, we implemented $S_4'$ instead of $S_4$ in our experiment,}
\end{figure}

If the parameter $\theta_2$ in matrix $S_4$ is 0, the matrices $D_2$ and $D_3$ both have a diagonal entry equal to 0 (Eq. S7), therefore the matrices $B$ and $C$ in Eq. S4 both have a singular value equal to 0, and one of the eigenvalues of matrix $BB^\dagger$ is 0. The determinant of matrix $BB^\dagger$ also equals to 0. Let
\begin{align*}
B = \left[\begin{matrix}
b_{11} & b_{12} \\
b_{21}& b_{22}
\end{matrix}\right],
\end{align*}
then
\begin{align*}
BB^\dagger = BB^\mathrm{T}  = \left[\begin{matrix}
b_{11} & b_{12} \\
b_{21} & b_{22}
\end{matrix}\right]\left[\begin{matrix}
b_{11} & b_{21} \\
b_{12} & b_{22}
\end{matrix}\right]
= \left[\begin{matrix}
b_{11}^2 + b_{12}^2 & b_{11}b_{21} + b_{12}b_{22} \\
b_{21}a_{11} + b_{22}a_{12} & b_{21}^2 + b_{22}^2
\end{matrix}\right]  = 0
\end{align*}
\begin{align*}
\Rightarrow b_{11}b_{22} - b_{21}b_{12} = 0 \Rightarrow |B|=0
\end{align*}

Similarly, we can get $|C|=0$. The specific CSD performed on $U_1$ and $U_2$ is shown as follows:
\begin{align}
U_1= \left[\begin{matrix}
-\frac{1}{2} & \frac{1}{2} & -\frac{1}{2} & \frac{1}{2}\\
\frac{1}{\sqrt{2}} & \frac{1}{\sqrt{2}} & 0 & 0\\
0 & 0 & \frac{1}{\sqrt{2}} & \frac{1}{\sqrt{2}}\\
-\frac{1}{2} & \frac{1}{2} & \frac{1}{2} & -\frac{1}{2}
\end{matrix}\right]  =
\left[\begin{matrix}
1 & 0 & 0 & 0\\
0 & 1 & 0 & 0\\
0 & 0 & 0 & 1\\
0 & 0 & -1 & 0
\end{matrix}\right]\left[\begin{matrix}
-\frac{1}{\sqrt{2}} & 0 & -\frac{1}{\sqrt{2}} & 0\\
0 & 1 & 0 & 0\\
\frac{1}{\sqrt{2}} & 0 & -\frac{1}{\sqrt{2}} & 0 \\
0 & 0 & 0 & 1
\end{matrix}\right] \left[\begin{matrix}
\frac{1}{\sqrt{2}} & -\frac{1}{\sqrt{2}} & 0 & 0\\
\frac{1}{\sqrt{2}} & \frac{1}{\sqrt{2}} & 0 & 0\\
0 & 0 & \frac{1}{\sqrt{2}} & -\frac{1}{\sqrt{2}}\\
0 & 0 & \frac{1}{\sqrt{2}} & \frac{1}{\sqrt{2}}
\end{matrix}\right]
%& = \left[\begin{matrix}
%\cos0\degree & -\sin0\degree & 0 & 0\\
%\sin0\degree & \cos0\degree & 0 & 0\\
%0 & 0 & \cos(-90\degree) & -\sin(-90\degree)\\
%0 & 0 & \sin(-90\degree) & \cos(-90\degree)
%\end{matrix}\right]\\
%&\left[\begin{matrix}
% \sin225\degree & 0 \cos225\degree & 0\\
% 0 & 1 & 0 & 0\\
% -\cos225\degree & 0 & \sin225\degree & 0\\
%0 & 0 & 0 & 1
%\end{matrix}\right]\\
%&\left[\begin{matrix}
%\cos45\degree & -\sin45\degree & 0 & 0\\
%\sin45\degree & \cos45\degree & 0 & 0\\
%0 & 0 & \cos45\degree & -\sin45\degree\\
%0 & 0 & \sin45\degree & \cos45\degree
%\end{matrix}\right]
\end{align}
\begin{align}
U_2= \left[\begin{matrix}
-\frac{1}{\sqrt{6}} & \frac{1}{\sqrt{2}} & -\frac{1}{\sqrt{7}} & \frac{2}{\sqrt{21}}\\
\frac{1}{\sqrt{6}} & \frac{1}{\sqrt{2}} & \frac{1}{\sqrt{7}} & -\frac{2}{\sqrt{21}}\\
0 & 0 & \frac{2}{\sqrt{7}} & \frac{3}{\sqrt{21}}\\
-\frac{2}{\sqrt{6}} & 0 & \frac{1}{\sqrt{7}} & -\frac{2}{\sqrt{21}}
\end{matrix}\right] =
\left[\begin{matrix}
\frac{1}{\sqrt{2}} & \frac{1}{\sqrt{2}} & 0 & 0\\
-\frac{1}{\sqrt{2}} & \frac{1}{\sqrt{2}} & 0 & 0\\
0 & 0 & 0 & 1\\
0 & 0 & -1 & 0
\end{matrix}\right]\left[\begin{matrix}
-\frac{1}{\sqrt{3}} & 0 & -\sqrt{\frac{2}{3}} & 0\\
0 & 1 & 0 & 0\\
\sqrt{\frac{2}{3}} & 0 & -\sqrt{\frac{1}{3}} & 0 \\
0 & 0 & 0 & 1
\end{matrix}\right]\left[\begin{matrix}
1 & 0 & 0 & 0\\
0 & 1 & 0 & 0\\
0 & 0 & \sqrt{\frac{3}{7}} & -\sqrt{\frac{4}{7}}\\
0 & 0 & \sqrt{\frac{4}{7}} & \sqrt{\frac{3}{7}}
\end{matrix}\right]
%& = \left[\begin{matrix}
%\cos(-45\degree) & -\sin(-45\degree) & 0 & 0\\
%\sin(-45\degree) & \cos(-45\degree) & 0 & 0\\
%0 & 0 & \cos(-90\degree) & -\sin(-90\degree)\\
%0 & 0 & \sin(-90\degree) & \cos(-90\degree)
%\end{matrix}\right]\\
%&\left[\begin{matrix}
% \sin215.3\degree & 0 \cos215.3\degree & 0\\
%0 & 1 & 0 & 0\\
%-\cos215.3\degree & 0 & \sin215.3\degree & 0\\
%0 & 0 & 0 & 1
%\end{matrix}\right]\\
%&\left[\begin{matrix}
%\cos0\degree & -\sin0\degree & 0 & 0\\
%\sin0\degree & \cos0\degree & 0 & 0\\
% 0 & 0 & \cos49.1\degree & -\sin49.1\degree\\
%0 & 0 & \sin49.1\degree & \cos49.1\degree
%\end{matrix}\right]
\end{align}

In this way, we simplify our optical circuitry and obtain the specific values of each HWP's angle in Fig. S6.

\subsection{Optical elements used in the experiment}

The beam displacers (BDs) we used are made from high optical grade calcites, and we used two sets of them in our experiment. One set has the size of $10\times10\times40 mm^3$, which separated the two beams of light by 4 mm; the other has the size of $15\times15\times20 mm^3$, which separated the two beams of light by 2 mm. Most of the HWPs we used are D-shaped zero-order half-wave plates at the wavelength of 810 nm, so that they can be easily added to the desired optical path without disturbing other paths, as shown in Fig. S7.
\begin{figure}[htp]
	\includegraphics[width=0.5\textwidth]{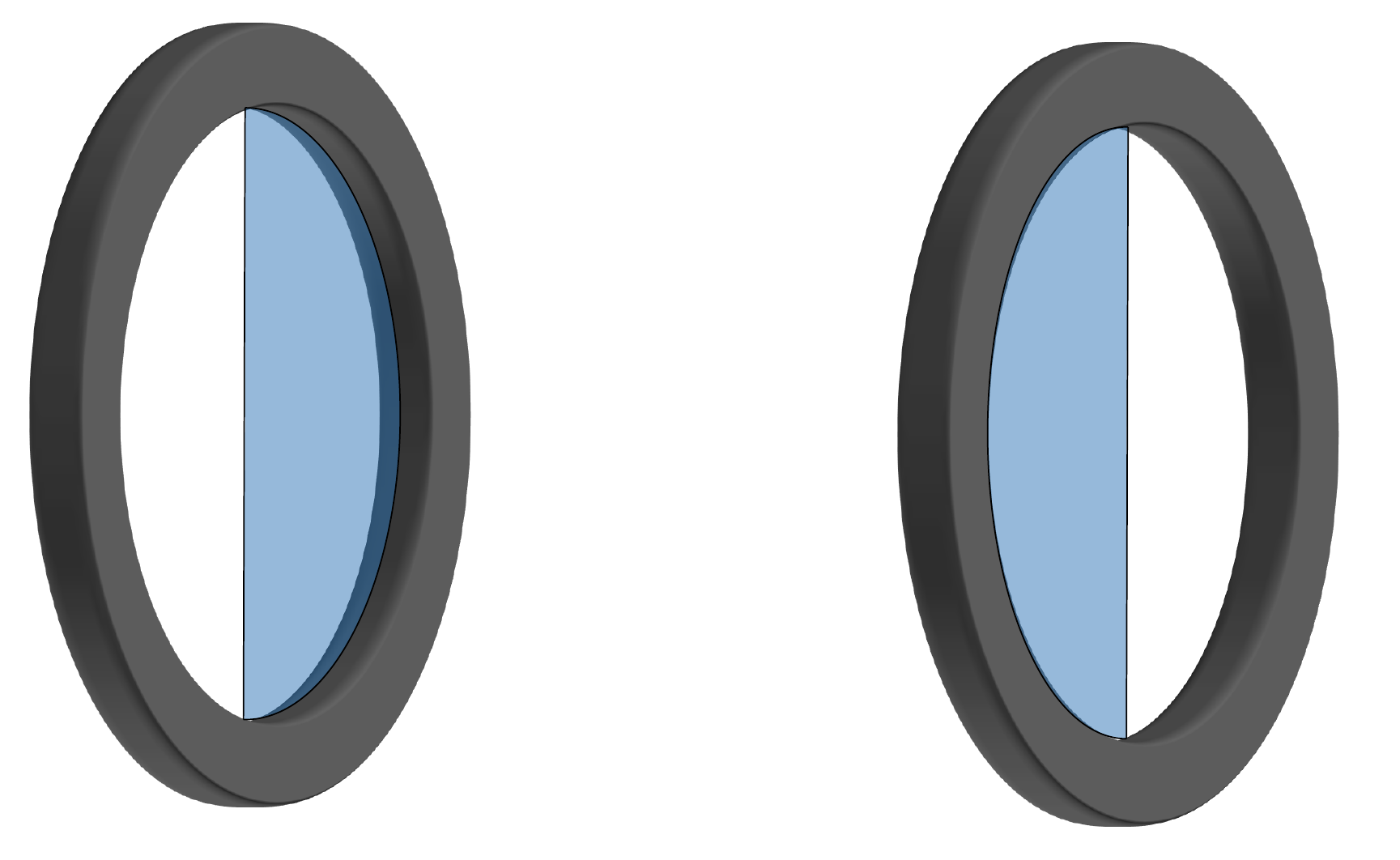}
	\caption{\label{fig:d_hwp} D-shaped half-wave plates we used in the experiment.}
\end{figure}

\subsection{Details of the experimental setup}

Based on the above algorithms and methods we can realize an arbitrary four-dimensional pseudo-Hermitian quantum walk. The module diagram is shown in Fig. S5. Based on the form of these matrices, we calculate the corresponding half-wave plate angles, as shown in Fig. S8\textbf{a}.
\begin{figure}[htp]
	\includegraphics[width=0.99\textwidth]{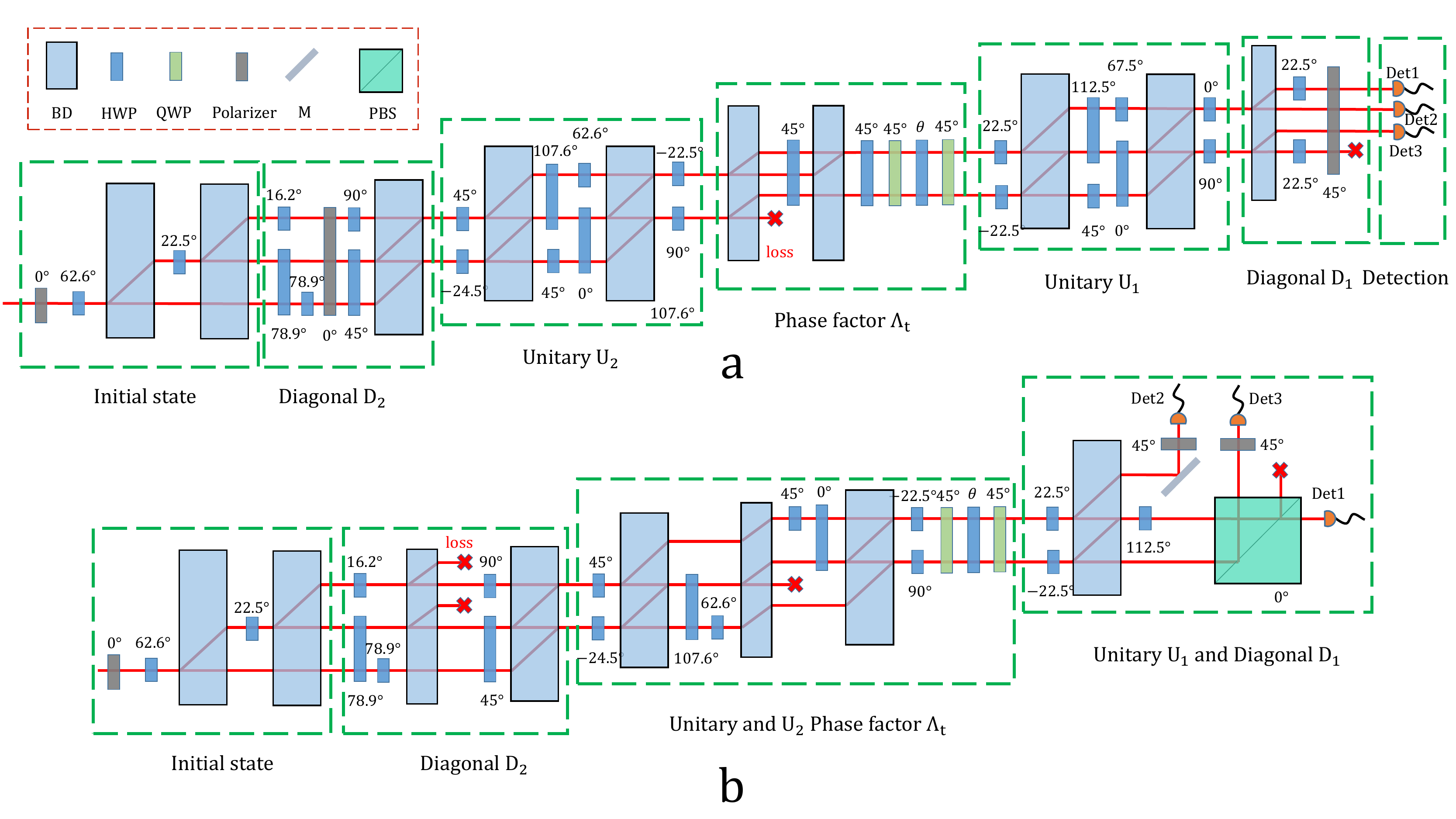}
	\caption{\label{fig_ setup}Detailed setup for realizing the three-vertex graph. \textbf{a}. Original optical circuitry with angles of wave plates. \textbf{b}. Simplified optical circuitry with angles of wave plates. Compared to the original setup, we implement the following changes: the polarizer in the diagonal matrix $D_2$ is replaced by a BD plus a blocker; the photons occupying mode $\ket{2V}$ are blocked; the last two BDs in the unitary matrix $U_1$ and the diagonal matrix $D_1$ are replaced by a PBS plus a mirror to separate photons occupying different modes.}
\end{figure}

We simplified the optical circuitry of Fig. S8\textbf{a} to the equivalent one shown in Fig. S8\textbf{b}. Firstly, the photons will be deflected slightly after passing through a polarizer which reduces the fiber coupling efficiency, so we replace the polarizer in the diagonal matrix $D_2$ of Fig. S8\textbf{a} with a BD and blockers, which have the equivalent function of allowing photons with specific polarization to pass through. Next is the unitary matrix $U_{2}$ and the phase factor matrix $\Lambda_{t}$. In Fig. S8\textbf{a}, four modes are separated after passing through the first BD in the phase factor matrix $\Lambda_{t}$, then the photons occupying mode $\ket{2V}$ are blocked. While in the simplified Fig. S8\textbf{b}, the photons occupying mode $\ket{2V}$ are blocked when implementing the unitary matrix $U_2$. With this equivalent module, we can save a BD and effectively shorten the length of the optical circuitry. In the unitary matrix $U_1$ of Fig. S8\textbf{a}, the two modes passing through the half-wave plates at $0\degree$ and $90\degree$, respectively, are orthogonal. In fact, the last two BDs in Fig. S8\textbf{a} make the three beams of light merge with and separate from each other again, so we can replace the last two BDs with a PBS and a mirror to separate the three modes.

\section{Details of implementing the four-dimensional quantum walk }
In this section, we show the details about realizing the pseudo-Hermitian CTQW on the four-vertex directed graph.

\subsection{Diagonalization of Hamiltonian}\label{sec:5a}
As shown in Eq.(7) in the main text, the Hamiltonian of our four-vertex graph is also a non-Hermitian matrix, whose diagonalization can be written as:
\begin{align}
H = \left[\begin{matrix}
1 & 0& -1 & -1\\
0&1 & -1 & -1\\
0 &-1& 2 & 0\\
-1&0&0&2
\end{matrix}\right]= \frac{1}{6} \left[ \begin{matrix}
2 & -1 & 1 & 0 \\
2 & -1 & -1 & 0 \\
1 & 1 & -1 & 1 \\
1 & 1 & 1 & -1\end{matrix}\right] \left[\begin{matrix}
0 & 0 & 0 & 0 \\
0 & 3 & 0 & 0 \\
0 & 0 & 1 & 0 \\
0 & 0 & 0 & 2\end{matrix}\right] \left[ \begin{matrix}
1 & 1 & 1 & 1 \\
-1 & -1 & 2 & 2 \\
3 & -3 & 0 & 0 \\
3 & -3 & 3 & -3\end{matrix}\right]= P \Lambda P^{-1}
\end{align}
and then we can obtain the time evolution operator:
\begin{align}\label{eq:evolution4}
e^{-iHt} = \frac{1}{6}e^{-1.5it} \left[ \begin{matrix}
2 & -1 & 1 & 0 \\
2 & -1 & -1 & 0 \\
1 & 1 & -1 & 1 \\
1 & 1 & 1 & -1\end{matrix}\right] \left[\begin{matrix}
e^{1.5it} & 0 & 0 & 0 \\
0 & e^{-1.5it} & 0 & 0 \\
0 & 0 & e^{0.5it} & 0 \\
0 & 0 & 0 & e^{-0.5it}\end{matrix}\right] \left[ \begin{matrix}
1 & 1 & 1 & 1 \\
-1 & -1 & 2 & 2 \\
3 & -3 & 0 & 0 \\
3 & -3 & 3 & -3\end{matrix}\right]
\end{align}
where the phase factor matrix can be achieved with the setup shown in Fig. S9. Here we only need to focus on the realization of the following matrices:
\begin{align}
P = \left[\begin{matrix}
2 & -1 & 1 & 0 \\
2 & -1 & -1 & 0 \\
1 & 1 & -1 & 1 \\
1 & 1 & 1 & -1\end{matrix}\right] \quad \textrm{and} \quad
P^{-1} =\left[\begin{matrix}
1 & 1 & 1 & 1 \\
-1 & -1 & 2 & 2 \\
3 & -3 & 0 & 0 \\
3 & -3 & 3 & -3\end{matrix}\right]
\end{align}
\begin{figure*}[htb]
	\includegraphics[width=0.35\textwidth]{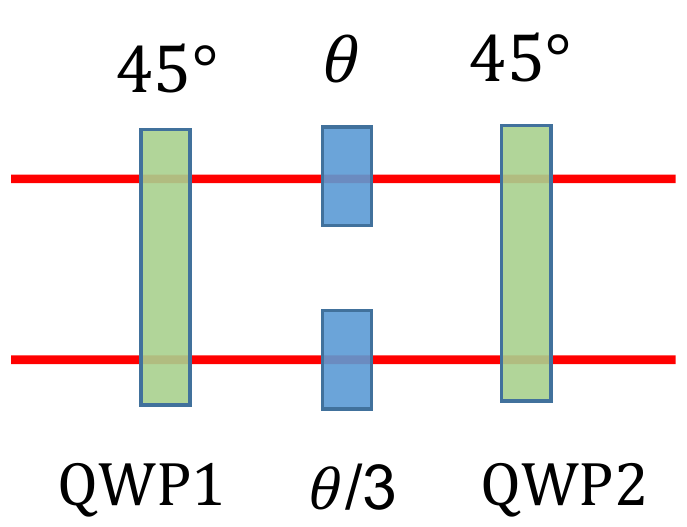}
	\caption{\label{fig_phase2}By using a combination of two quarter-wave plates (oriented to 45\degree) and two half-wave plates (oriented to $\theta$ and $\theta$/3), we can realize the phase factor matrix of our four-dimensional pseudo-Hermitian CTQW.}
\end{figure*}

\subsection{Singular value decomposition}
According to the method introduced in Sec. \uppercase\expandafter{\romannumeral2}, we perform SVD on the non-unitary $P$ and $P^{-1}$. In order to facilitate the experiment we let SVD produce a unitary block-diagonal matrix $V_1$, so that we can realize it by using only half-wave plates. Therefore, we decompose the matrices as follows:
\begin{align*}
P = 3.2566 &  \left[ \begin{array}{cccc}{-0.6768} & {-0.2049} & {0.3717} & {-0.6015} \\ {-0.6768} & {-0.2049} & {-0.3717} & {0.6015} \\ {-0.2049} & {0.6768} & {-0.6015} & {-0.3717} \\ {-0.2049} & {0.6768} & {0.6015} & {0.3717}\end{array}\right]   \left[ \begin{array}{ccccc}{1} & {0} & {0} & {0} \\ {0} & {0.5657} & {0} & {0} \\ {0} & {0} & {0.7026} & {0} \\ {0} & {0} & {0} & {0.2684}\end{array}\right] \times \\
& \left[ \begin{array}{cccc}{-0.9571} & {0.2898} & {0} & {0} \\ {0.2898} & {0.9571} & {0} & {0} \\ {0} & {0} & {0.8507} & {-0.5257} \\ {0} & {0} & {-0.5257} & {-0.8507}\end{array}\right] = 3.2566 \times U_1 D_1 V_1
\end{align*}
where $U_1$ is a unitary matrix and $D_1$ is a diagonal matrix. Note that all the singular values are positive real numbers. Since the realization of some two-dimentional real unitary matrices requires more than one half-wave plate, we need to change the sign of some rows and columns of the unitary matrix into the form which can be realized using only one HWP. Note that changing the sign of entries of a diagonal matrix won't effect the complexity of our optical circuitry. Correspondingly, we change the form of $U_1$ and $D_1$ in the following:
\begin{align*}
U_1 D_1 V_1 = & \left[ \begin{array}{cccc}{0.6768} & {0.2049} & {-0.3717} & {-0.6015} \\ {0.6768} & {0.2049} & {0.3717} & {0.6015} \\ {0.2049} & {-0.6768} & {0.6015} & {-0.3717} \\ {0.2049} & {-0.6768} & {-0.6015} & {0.3717}\end{array}\right]   \left[ \begin{array}{ccccc}{1} & {0} & {0} & {0} \\ {0} & {0.5657} & {0} & {0} \\ {0} & {0} & {0.7026} & {0} \\ {0} & {0} & {0} & {-0.2684}\end{array}\right] \times \\
& \left[ \begin{array}{cccc}{-0.9571} & {0.2898} & {0} & {0} \\ {0.2898} & {0.9571} & {0} & {0} \\ {0} & {0} & {0.8507} & {-0.5257} \\ {0} & {0} & {-0.5257} & {-0.8507}\end{array}\right]
\end{align*}
Similarly, we decompose $P^{-1}$ as follows:
\begin{align*}
P^{-1} = 6.8644 &  \left[ \begin{array}{cccc}{-0.9571} & {-0.2898} & {0} & {0} \\ {0.2898} & {-0.9571} & {0} & {0} \\ {0} & {0} & {0.8507} & {0.5257} \\ {0} & {0} & {-0.5257} & {0.8507}\end{array}\right]   \left[ \begin{array}{ccccc}{0.2684} & {0} & {0} & {0} \\ {0} & {0.4745} & {0} & {0} \\ {0} & {0} & {0.3820} & {0} \\ {0} & {0} & {0} & {-1}\end{array}\right] \times \\
& \left[ \begin{array}{cccc}{-0.6768} & {-0.6768} & {-0.2049} & {-0.2049} \\ {0.2049} & {0.2049} & {-0.6768} & {-0.6768} \\ {0.3717} & {-0.3717} & {-0.6015} & {0.6015} \\ {-0.6015} & {0.6015} & {-0.3717} & {0.3717}\end{array}\right] = 6.8644 \times V_2 D_2 U_2
\end{align*}

The realization of unitary block-diagonal matrices $V_1$ and $V_2$ requires only two and three HWPs, respectively. The diagonal matrices $D_1$ and $D_2$ can be achieved by the optical path of Fig. S10, and the only thing we need is to calculate the angles of those HWPs. Next we will introduce the method of implementing the unitary matrices $U_1$ and $U_2$.

\subsection{Cosine-Sine decomposition of $U_1$ and $U_2$}
The specific CSD performed on $U_1$ and $U_2$ is shown as follows:
\begin{align*}
U_1=\left[ \begin{array}{cccc}{-0.7071} & {-0.7071} & {0} & {0} \\ {0.7071} & {-0.7071} & {0} & {0} \\ {0} & {0} & {0.7071} & {-0.7071} \\ {0} & {0} & {0.7071} & {0.7071}\end{array}\right]\left[ \begin{array}{cccc}{0} & {0} & {1} & {0} \\ {0} & {1} & {0} & {0} \\ {-1} & {0} & {0} & {0} \\ {0} & {0} & {0} & {1}\end{array}\right] \left[ \begin{array}{cccc}{-0.2898} & {0.9571} & {0} & {0} \\ {-0.9571} & {-0.2898} & {0} & {0} \\ {0} & {0} & {0.5257} & {0.8507} \\ {0} & {0} & {-0.8507} & {0.5257}\end{array}\right]
\end{align*}
\begin{align*}
U_2=\left[\begin{array}{cccc}{0.2898} & {0.9571} & {0} & {0} \\ {0.9571} & {-0.2898} & {0} & {0} \\ {0} & {0} & {-0.5257} & {0.8507} \\ {0} & {0} & {0.8507} & {0.5257}\end{array}\right]\left[ \begin{array}{cccc}{0} & {0} & {-1} & {0} \\ {0} & {1} & {0} & {0} \\ {1} & {0} & {0} & {0} \\ {0} & {0} & {0} & {1}\end{array}\right] \left[\begin{array}{cccc}{-0.7071} & {0.7071} & {0} & {0} \\ {-0.7071} & {-0.7071} & {0} & {0} \\ {0} & {0} & {0.7071} & {0.7071} \\ {0} & {0} & {-0.7071} & {0.7071}\end{array}\right]
\end{align*}
According to the setup shown in Fig. S6 and the result of CSD algorithm, we can calculate the specific values of each HWP's angle, and finally implement the unitary matrices $U_1$ and $U_2$.

\subsection{Optical elements used in the experiment}
Besides the form shown in Eq. S10, a real four-dimensional diagonal matrix can also be written as:
\begin{align}\label{eq:diagonal2}
D = \lambda\left[\begin{matrix}
\sin2\theta_1 & & & \\
& \sin2\theta_2 & &\\
& & \sin2\theta_3 & \\
& & & \sin2\theta_4
\end{matrix}\right]
\end{align}
We can realize it by tuning the parameters $\theta_1, \theta_2, \theta_3, \theta_4$ as shown in Fig. S10\textbf{a}.

According to Fig. S10\textbf{a}, the implementation of diagonal matrices $D_1$ and $D_2$ needs four HWPs with different angles, respectively. In order to shorten the optical circuitry and keep the four optical paths from disturbing each other, we use a composite module made by four HWPs with different orientation angles of their fast axis. Then we make the four beams of light pass through each of them, respectively, as shown in Fig. S10\textbf{b}.
\begin{figure*}[htp]
	\includegraphics[width=0.99\textwidth]{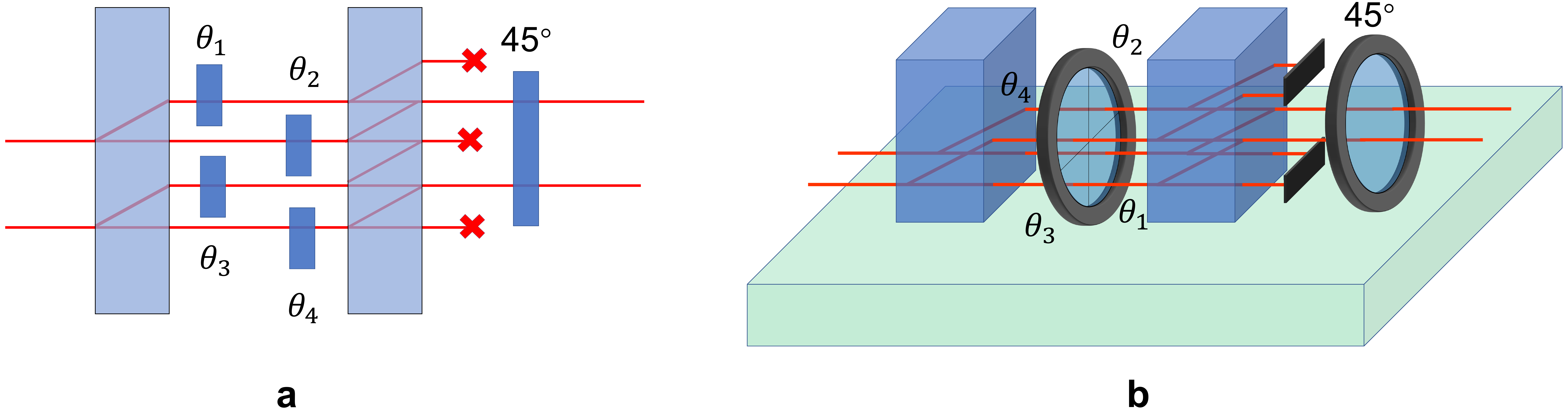}
	\caption{\label{fig:m_hwp}.\textbf{a}. The general diagram for realizing a four-dimensional real diagonal matrix. \textbf{b}. The composite-HWP module we used for realizing the four-dimensional diagonal matrices.}
\end{figure*}

\subsection{Details of the experimental setup}
Based on the above algorithms and methods, we can calculate the corresponding angle of each half-wave plate, as shown in Fig. S11.

We simplified the optical circuitry of Fig. S11\textbf{a} to the equivalent one shown in Fig. S11\textbf{b}. Firstly, the two adjacent HWPs oriented to 45 \degree while implementing the matrices $D_2$ and $V_2$ can be simplified to a equivalent unit matrix, so that we can remove the two HWPs. Besides, the two adjacent HWPs oriented to 45 \degree while implementing the matrices $D_1$ and $U_1$ can be simplified to a equivalent unit matrix, so that we can also remove the two HWPs to shorten the optical circuitry.
\begin{figure*}[htb]
	\includegraphics[width=0.99\textwidth]{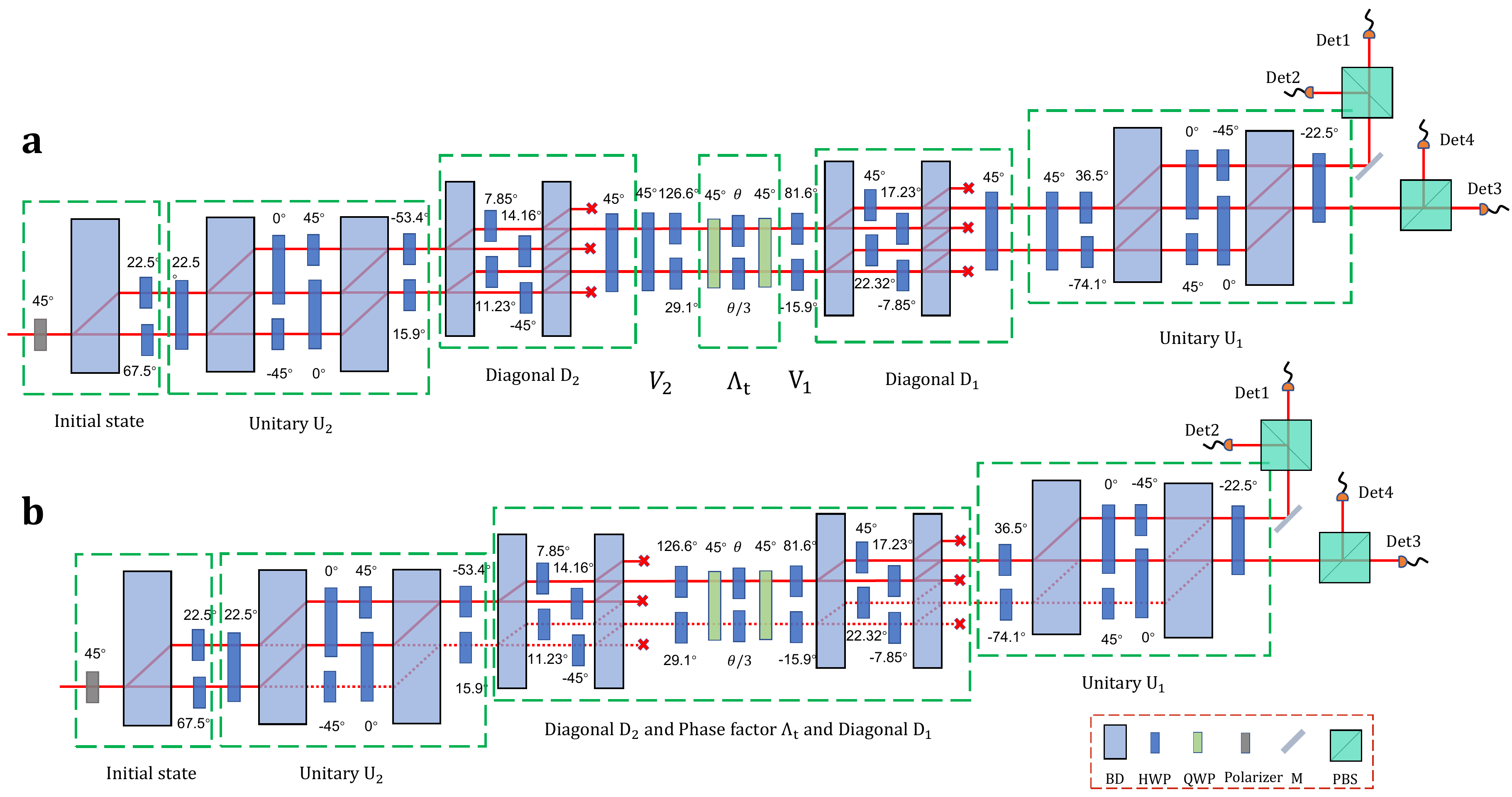}
	\caption{\label{fig:setup_4}Detailed setup of our experiment. \textbf{a}. Original optical circuitry with angles of wave plates. \textbf{b}. Simplified optical circuitry with angles of wave plates. Compared to the original setup, we remove four HWPs so that the optical path is shortened.}
\end{figure*}

\end{document}